\documentclass[12pt]{article}
\usepackage{a4wide, amsmath, latexsym, epsfig,amsfonts,
  psfrag,graphicx, rotating, fancyhdr, tabularx, afterpage,booktabs}
\usepackage{subfigure}
\usepackage{amssymb}
\usepackage{setspace}

\newcommand\ba{\begin{eqnarray}}
\newcommand\ea{\end{eqnarray}}

\begin{document}
\begin{titlepage}
 
\begin{flushright} 
{ IFJPAN-IV-2014-9} 
\end{flushright}

\vspace{0.2cm} 
\begin{center}

{\Huge \bf  TauSpinner: a tool for simulating CP effects in $ H \to \tau \tau$ decays at LHC }
\end{center}
\vspace*{5mm}

\begin{center}
   {\bf 
        T. Przedzi\'nski$^{a}$, E. Richter-W\c{a}s$^{b}$
 and Z. W\c{a}s$^{c}$  }\\
       {\em $^a$ The Faculty of Physics, Astronomy and Applied Computer Science, \\ 
Jagellonian University, Reymonta 4, 30-059 Cracow, Poland}\\
       {\em $^b$ Institute of Physics, \\ 
Jagellonian University, Reymonta 4, 30-059 Cracow, Poland}\\
       {\em $^c$  Institute of Nuclear Physics, PAN,
        Krak\'ow, ul. Radzikowskiego 152, Poland} 

\end{center}
\vspace{.1 cm}
\begin{center}
{\bf   ABSTRACT  }
\end{center} 

In this paper,
we discuss application of the  {\tt TauSpinner} package as a simulation tool
for  measuring the CP state of the newly discovered Higgs boson
using the transverse spin correlations in the $H \to \tau \tau$ decay channel.
We discuss application for its main background $Z/\gamma^* \to \tau \tau$ as well. 
The  {\tt TauSpinner}  package allows one to add, with the help of weights, 
transverse spin correlations 
corresponding to any mixture of scalar/pseudoscalar state,
on already existing events using information from the kinematics 
of outgoing $\tau$ leptons and their decay products only. 
This procedure can be used  when polarimetric vectors of the $\tau$s decays and 
density matrix for $\tau$-pair production are not stored with 
the  event sample.

We concentrate on the well-defined  
effects for the Higgs (or Higgs-like scalar) decays, which are physically separated from the
production processes. 
%Parity effects in case of distinct nature of the resonance Higgs production mechanisms of possible Standard Model  
%extensions will be left for the forthcoming paper.
{\tt TauSpinner}  also
allows to reintroduce (or remove) spin correlations  to events from Drell-Yan 
$Z/\gamma^* \to \tau \tau$ process, the main background for the Higgs parity 
observables, again with the help of weights only. 

From the literature, we recall well-established observables, developed for 
measuring the CP of the Higgs,
 and use them as benchmarks for illustrating applications 
of the {\tt TauSpinner} package.  
We also include a description of the code and prepared testing examples.

\vskip 1 cm

%Plots  ported to the text of the paper: {\bf \today}. 

\vspace{0.2 cm}
%\centerline{ \bf DRAFT \today }
 
 \vspace{1cm}
\begin{flushleft}
{   IFJPAN-IV-2014-9\\
 June, 2014}
\end{flushleft}
 
%%%%%%%%%%%%%%%%%%%%%%%%%%%%%%%%%%%%%%%%%%%%%%%%%%%%%%
\vspace*{1mm}
\bigskip
%%%%\vfill
%\footnoterule
\noindent
{\footnotesize \noindent% 
%Em
}
\end{titlepage}

\section{Introduction}

In the year 2012, ATLAS and CMS Collaborations published the discovery of a new resonance 
\cite{AtlasHiggs2012, CMSHiggs2012} in the search of the Standard Model Higgs boson H \cite{Englert64, Higgs64}, 
with a mass of about 125~GeV. The next experimental challenge soon became to measure and compare
its properties with the Standard Model predictions. In 2013 the ATLAS Collaboration reported 
\cite{AtlasScalar2013} that the data are compatible with the Standard Model $J^{P} = 0^{+}$ 
quantum numbers for the Higgs boson, whereas all alternative hypotheses studied 
are excluded at the confidence level above 97.8\%. 

The Higgs boson decay to $\tau$ pairs has been directly confirmed so far only as an evidence, 
with observed signal significance on the level of 4.2$\sigma$ (ATLAS, \cite{ATLASHtautau}) and 3.2~$\sigma$ (CMS, \cite{CMSHTautau}).
The run II of LHC, which is starting next year, will hopefully bring observation in this channel also above 5$\sigma$.    

Measurement of the Higgs CP state in the decay channel to $\tau$'s will be an 
avenue of the physics programme for measuring Higgs properties in its decays to {\it fermions}. Tau leptons offer a unique
opportunity of being excellent polarimeter probes, as the spin correlations, both longitudinal 
and transverse, are propagated to their measurable decays products. 
Availability of the multi-body decays: to single $\pi^\pm$ or 3 $\pi^\pm$, 
and via resonances like $\rho\to \pi^\pm \pi^0$,  allows to study and explore correlations between 
decay products. 

In  this paper we first recall sensitive observables for the measurement of the
CP Higgs boson state in the decay to $\tau$ leptons and present new developments in the 
{\tt TauSpinner} package which can be used for studying such observables on 
 Monte Carlo samples or data embedded events with the help of weights with easy to configure options. Both longitudinal and 
 transverse spin 
correlations are now available in {\tt TauSpinner} \cite{Czyczula:2012ny,Banerjee:2012ez, TauSpinner2014} 
and can be modeled with this package 
starting from a sample which does not included such effects.  This implementation allows for simulation 
of transverse spin effects for the mixture of scalar/pseudoscalar Higgs boson state and also for 
the main background, the Drell-Yan (DY) process $Z/\gamma^* \to \tau \tau$. 
All with the help of the weights calculated after Monte Carlo samples are already generated. In particular there is no need for polarimetric vectors for $\tau^\pm $ decays and density matrices for $\tau$-pair 
production to be stored in the event sample.  For some generators 
such information may be made available,  
but not  in the case of embedded $\tau$'s, see e.g.~\cite{ATLASHtautau}. 

Because correlations are introduced  by spin weights $wt$, the kinematical configuration of each event remains
intact,  
 statistical fluctuations on the $(wt-1)$ difference only 
introduce errors to the results of  
the comparisons.  Weighted and unweighted samples are correlated.  
Use of weighted samples allows  to study different theory models
without the need of time-consuming simulation of  the
detector responses: for arbitrary 
choice of  Higgs parity state assumed for its decay and/or for its production matrix element.
In the present paper we  concentrate on Higgs decay. For scalars, decays are independent from the
kinematic configuration of  production process which can be treated separately. 
%{\it One should stress that at this step we concentrate on the Higgs decay only: NIE ROZUMIEM.
%Czy chodzi o to ze mozna tylko symulowac dowolna mixcture scalar/pseudoscalar? Jezeli tak to napisz wprost.
%Albo moze masz wagi takze dla spin=2 tylko nie poprawiasz correlacji w mechanizmie produkcji?} 

As we will show, effects from the transverse spin correlations in decays are rather small. 
Let us note however that the data analysis techniques which have been developed during Run~I of LHC
have reached unprecedented level of sophistication, both on the statistical treatment and events 
classification techniques. As a result, it is  impossible to judge the feasibility  to measure the 
Higgs boson CP with the $\tau\tau$ decay channel. We do not address this question in our paper.
But having a program available  to properly model such effects  will certainly be a key to the success 
of such difficult measurement.    

Our paper is organized as follows; in the following Section~\ref{sec:transverse},
we briefly recall the definition of the longitudinal and transverse spin correlations.
In  Section \ref{sec:CPsensitive},  we recall definition of the 
spin-sensitive observables in the $H \to \tau \tau$ decay.  In  Section \ref{sec:Newdevelopment}, we 
present new development of their implementation in  {\tt TauSpinner} package. The next 
Sections \ref{sec:Case125} and \ref{sec:CaseDY}, for signal and background respectively,
 collect some benchmark numerical results obtained with {\tt TauSpinner}
package and discuss possible improvements in the definition of those observables. 
Appendix \ref{sec:technical}  gives more technical details concerning the usage of  {\tt TauSpinner}. 
Several tests are also included to facilitate usage of the package and checks on its installation. 
Summary, Section \ref{sec:summary}, closes the paper.

\section {Transverse spin correlations }
\label{sec:transverse}

Spin correlations of $\tau \tau$ pairs from Higgs boson decays are sensitive to its parity~\cite{Nelson89}.
This sensitivity is reflected in angular correlations of secondary decay products, in particular in 
the acollinearity distributions of the $\pi^+ \pi^-$ from $\tau^+\tau^-$ decays. 
The spin density matrix for the two $\tau$s resulting from the decay of the state which is a mixture 
of scalar/pseudoscalar is given, using convention of \cite{Kramer:1994jn}, by the formula below
\begin{equation}  
\Gamma(H_{mix}\to \tau^{+}\tau^{-}) \sim 1-s^{\tau^{+}}_{\parallel}
s^{\tau^{-}}_{\parallel}+ s^{\tau^{+}}_{\perp}
R(2\theta)~s^{\tau^{-}}_{\perp},
\label{densi}  
\end{equation}   
where $R(2\theta)$ can be understood as  an operator for the rotation by
an angle $2\theta$ (double the mixing scalar-pseudoscalar angle $\theta$)   
around the ${\parallel}$ direction, i.e. of $\tau^+ \tau^-$ momenta in H rest-frame.  
The $s^{\tau^{-}}$ and $s^{\tau^{+}}$ are  the $\tau^\pm$ polarization
vectors, which are defined  in their respective rest frames\footnote{
Note the distinct frame conventions used in {\tt TauSpinner} with respect
to other works, e.g. Ref.~\cite{Desch:2003rw}.}.

The $R(2\theta)$ spin density matrix, effectively a rotation matrix,
reduces to just  a diagonal one for the pure parity state: $R_{xx}=R_{yy}=1$ 
for scalar and $R_{xx}=R_{yy}=-1$ for pseudoscalar.

\section {CP sensitive observables}
\label{sec:CPsensitive}

It is clear from formula (\ref{densi}), that to measure effects due to 
Higgs parity, it is necessary to understand how 
effects due to the $s^{\tau^{\pm}}_{\perp}$ component of the polarization translate
to {\it measurable} quantities.

The $\tau$ lepton is an unstable fermion often decaying hadronically. 
Such process was described for the first time in \cite{Tsai:1971vv},  its
differential decay width is given by the formula
\begin{equation}\label{eq:3}
\frac{d\sigma}{d\Omega_{decay}}=\frac{d\sigma^{no polarized}}{d\Omega_{decay}}(1+{s}\cdot{h}) 
\end{equation}

where $h$ depends on the actual decay mode and its kinematical
configuration, the $\frac{d\sigma^{no polarized}}{d\Omega_{decay}}$ denotes
differential decay width in case when polarization is absent, and a vector $s$ (conveniently 
defined in its rest-frame), parametrizes its spin density
matrix.
In case of $\tau^\pm \to \pi^\pm \nu_\tau$ the $h$ 
is defined by the direction of $\nu_\tau$ momentum (in $\tau$ rest-frame). 
Also for other 
$\tau$ decay modes $h$ is correlated with the  $\nu_\tau$ momentum.
The key to the measurement, used in all approaches, in particular in
\cite{Kramer:1994jn,Desch:2003rw,Bower:2002zx,Desch:2003mw,Rouge:2005iy},
is to have at least partial control over the
relative orientation of the  planes spanned by the momenta of $\tau^\pm$ 
and its decay product $\nu_\tau$ (in the Higss boson rest-frame).
Unfortunately, all these quantities are at best difficult to measure and depending
on the detector conditions one of the choices may be better than other ones. 
The actual choice of the observable among those proposed in these references
will of course depend  on details of the  detector and background understanding.
Further improvements with respect to these references can be expected, similarly as it was
the case  at a time of LEP measurements, see e.g.~\cite{Davier:1992nw} for the concept of optimal variables.

Thanks to the prolonged effort, especially in the context of preparing physics programme 
of the future linear collider, observables sensitive to the CP spin effects were designed for 
final states involving $\tau$ leptons. We would like to recall here two such observables, 
well established in the literature. 

In Ref.~\cite{Kramer:1994jn} it was shown that for the acollinearity angle defined in the rest frame 
of the Higgs boson,  the shape at the end-point of the acollinearlity 
distributions for the two pions resulting from Higgs decay can be easily interpreted as
a consequence of spin parity. Reconstruction of the $\tau$ 4-momentum could be avoided,
but precise control of the Higgs rest-frame was necessary. 
It is rather easy to understand that indeed the 
transverse spin correlations, carrying information on Higgs parity, 
can be translated into observable such as acollinearity in $H \to \tau^+ \tau^-$, 
$\tau^\pm \to \pi^\pm \nu$ decay chain.  

In Ref.~\cite{Bower:2002zx} an alternative observable was proposed for 
 the $H \to \tau^+ \tau^-$, $\tau^\pm \to \rho^\pm \nu$,
$\rho^\pm \to \pi^\pm \pi^0$ decay chain. It was suggested to abandon reconstruction of the 
Higgs rest frame and $\tau$ four-momenta, instead use the $\rho^+ \rho^-$ rest frame which has the advantage
that it is built only from directly visible decay products of the  $\rho^+$ and $ \rho^-$. 
Reconstruction of the $\tau^+$ and  $\tau^-$ rest frames or momenta was not necessary.

One-dimensional angular distribution of the acoplanarity angle $\theta_{acop}$ 
between $\pi^+ \pi^0$  and $\pi^- \pi^0$ decay planes in the $\rho^+ \rho^-$ pair rest 
frame was proposed as CP sensitive observable. To establish this sensitivity, one has to define two categories
of events depending on the sign of the energy difference between $\pi^\pm$ and $ \pi^0$.
More specifically  the energy differences are defined as follows: 
 $y_+$  of $\pi^+ \pi^0$ from $\rho^+$, and $y_-$ 
of $\pi^- \pi^0$ from $\rho^-$ and categories depend on sign of the $y_+ \cdot y_- $  product.
The distribution of  the acoplanarity angle $\theta_{acop}$  
in each category separately is sensitive to the CP of the Higgs boson. 
%The  $y_+ y_- $ selection is used to correlate the decay plane of $\rho^\pm $ 
Energy differences,  $y_\pm$,  measured in the laboratory frame can be used. 
For better sensitivity of the observable, one can attempt to reconstruct those quantities in 
 the $\tau$'s rest frames. The second option is however more challenging experimentally. Also
the  $\tau^\pm \to a_1^\pm \nu$ decay chain is potentially very interesting, however 
in this decay chain it is much more difficult to construct CP-sensitive observable that is easy to interpret and use.

The effects of detector smearing for the measurements of $\pi$'s directions
and energies were studied in~\cite{Bower:2002zx}. The sensitivity to parity of the Higgs boson was preserved
even if relatively large smearing was allowed, as long as the relative orientation
of  $\pi^0$'s direction with respect to $ \pi^\pm$ could be established.
Precision for the measurement of energies was of no concern. Even large smearing
was found not to deteriorate the CP sensitivity of the observable.

\section {New development of the {\tt TauSpinner} code}
\label{sec:Newdevelopment}

The {\tt TauSpinner} package  \cite{Czyczula:2012ny,Banerjee:2012ez, TauSpinner2014} represents 
a tool which can be used to modify $\tau$ spin effects in samples where intermediate states
decay into final states including  $\tau$ leptons. 
As an input, generated samples of events featuring $\tau$ leptons produced from intermediate state 
$W$, $Z$, Higgs bosons are used. 
The information on the polarization and spin correlations is reconstructed from the kinematics of the 
$\tau$ leptons (also $\nu_\tau$ in case of $W$ mediated processes) and $\tau$ decay products alone. 
The hard process parton level scattering flavour/kinematical configurations, in particular $x_1$, $x_2$ for PDFs, are reconstructed
from the known centre-off-mass energy,  rapidity and virtuality of the $\tau^+\tau^-$ pair. The convolution of the 
parton density functions (PDF) at given $x_1$, $x_2$ and the effective Born level cross sections of the hard process
can be used for  calculating respective spin weights. Depending on whether
longitudinal and transverse spin correlations were included  in the original sample, the {\tt TauSpinner} package calculates
the corresponding  weight, on the event-by-event basis to model the required spin effects. 
It allows to model spin correlations of the scalar
or pseudoscalar Higgs boson or desired mixture of both states, starting from the same sample of generated events. 
Longitudinal spin correlations are 
modeled properly for the vector boson  $W$, $Z$ intermediate state as well. As discussed below, the first 
implementation  for modeling the transverse spin correlations in the $Z$ intermediate state
is also prepared.

By calculating spin weights, attributed on the event-by-event basis, {\tt TauSpinner} enables numerical evaluation of 
the spin effects on experimentally measured distributions and/or modification of the spin effects.

Because of the modularity of the design, most of the algorithms and solutions used in {\tt Tauola universal interface} package \cite{Golonka:2003xt}  
could have been used in {\tt TauSpinner}.  
In {\tt Tauola universal interface} 
longitudinal and transverse spin correlations are implemented as part of event generation since a long time. 
Spin effects  originate directly from the properties of  the matrix elements used. 
There are necessary extensions and new conditions, for algorithms of {\tt Tauola universal interface}
(and later {\tt Tauola++ universal interface}  \cite{Davidson:2010rw} as well)
to be useful for  inserting or modifying spin effects {\it post-fact} on already stored events.
 The numerical 
stability of the reweighting procedure (one has to boost back $\tau$ decay products from laboratory frame to $\tau$ rest frames)
has to be assured.  Also
the lack of information  available in the event record\footnote{In particular for event build from embedded data.},
necessary for calculating  the weights such as the flavours of incoming quarks which are not known, has to be overcome. 
To construct the missing information for the kinematics and/or flavour of hard process partons
an additional  algorithm, at present based on leading-log approximation only,  had to be introduced.
This is discussed in some more details in \cite{TauSpinner2014} and in other papers discussing {\tt TauSpinner} applications 
\cite{Czyczula:2012ny,Banerjee:2012ez}. 
To evaluate technical correctness of those algorithms it was sufficient to compare, for a given analysis selection,  
results from samples generated 
using the {\tt Tauola++ universal interface} and including all effects from the beginning, with the ones where effects were
obtained using the {\tt TauSpinner} package. Further work,  based on calculations with  approximations beyond
leading logarithm, will  allow to improve the precision of both the {\tt TauSpinner} and the 
{\tt Tauola++ universal interface} packages. This can not be completed on the basis of programs featuring leading logarithm 
approximations, one may need to wait until full NLL Monte Carlo programs become available \cite{Jadach:2012vs}.
At present comparisons with fixed order results such as \cite{vanHameren:2008dy}  or with NLO 
simulations in particular \cite{Alioli:2008tz,Nason:2009ai} are planned.

\subsection{Transverse spin correlations in $H \to \tau \tau$ decay}
The effects of transverse spin correlations resulting in sensitivity to Higgs parity in case of 
decays to $\tau$ leptons was already installed in {\tt Tauola universal interface}~\cite{Golonka:2003xt} some time ago.
 
We have discussed Higgs boson CP  observability 
already  in \cite{Bower:2002zx}. The method was later extended to the mixed parity case \cite{Desch:2003rw}. 
The $\tau$ decays to two pion final states were used to present numerical results.
The transverse spin correlations
manifest themselves in the form of a directly
observable acoplanarity\footnote{
This was contrary to  work \cite{Kramer:1994jn}
where difficult to measure acollinearity  angle between $\pi^+$ and $\pi^-$ directions (in the rest-frame of Higgs)
was discussed. Both $\tau$'s were to decay to single pion states.}  angle between (oriented)-decay planes of $\rho^+$ and $\rho^-$. %TP: is (oriented)-decay correct?
This observable was one of the  basic benchmarks for  implementing  the transverse spin
in   {\tt Tauola++ universal interface}  \cite{Davidson:2010rw}, the 
direct continuation of ideas presented in \cite{Golonka:2003xt}.

The application of the {\tt TauSpinner} package for simulating effects of the longitudinal spin correlations 
has been discussed in \cite{TauSpinner2014}. Presently, important elements of the algorithms implementing transverse  
spin correlations have  been ported  to this package. With the method of events reweighting, 
the algorithms are able to reintroduce
correlations to the kinematics of already generated  $H \to \tau \tau$ events without those effects. 
This is a useful technique because it allows one to work on  generated Monte Carlo events and also on 
the data embedded samples. 

\subsection{Transverse spin correlations in  $Z \to \tau \tau$ decay}
From now on, transverse spin correlations are available with {\tt TauSpinner}, for the  DY process, $Z/\gamma^* \to \tau\tau$. 
The spin correlation matrix $R_{ij}$ is calculated for {\tt Tauola++ universal interface} by the electroweak library 
\cite{Andonov:2008ga}, at one loop precision level. 
We have checked that,  
the  non diagonal part  of the spin density matrix remains close to zero 
in high energy limit, as expected from analytic form of \cite{jadach-was:1984} applicable for low virtualities of $\tau\tau$ pairs. 
The  $R_{xx}$, and $ R_{yy}$ components of $\tau^+ \tau^-$ pair spin density matrix (already 
precalculated and stored for {\tt Tauola++ universal interface} in text files)  are made available for the
use of {\tt TauSpinner}.  As the actual calculation of spin weight was identical, and only parton level kinematical 
configuration was obtained differently, it was technically rather simple. 

Let us elaborate a bit more on the actual implementations of the spin correlations.

One can represent the complete  differential cross section for 
$\tau$ pair production and decay as the combination of several segments. 
The separation into $\tau$ decays and production components is exact, thanks to the narrow width of the $\tau$.
Both phase space and matrix elements can be factorized into the part corresponding to production and the 
one corresponding to the decays. 
That is the property which is used in {\tt Tauola} since the beginning \cite{Jadach:1990mz}. For the
cross section calculation, separation into  production process and the decay is not complete, it requires introducing 
the spin weight $wt$. Only this weight depends simultaneously on kinematical
variables of the production and of the $\tau$ decays.  This dependence is regular and consists of 
contraction of 4$\times4$ density matrix $R_{ij}$ for $\tau$ pair production and polarimetric 
vectors $h_i^+$, $h_j^-$ for each $\tau^\pm$ decay. The following properties hold by construction:  $0<wt<4$  and  $<wt>=1$.
This is  used in 
\cite{Jadach:1999vf,Jadach:1993yv} and does not require any approximation. It is used by the
{\tt Tauola universal interface} and {\tt TauSpinner} as well. 
The problem maybe be simplified however; depending on the approximations some 
of the components of $R_{ij}$ may be ignored and set to zero. Also, if only collinear helicity-like degrees are 
taken into account, detailed definition of frames used for $\tau$'s quantization is not necessary\footnote{
That simplifies the problem, there is no need for control on relative orientation of all quantization frames. 
The question of the choice of quantization frames  has been discussed in detail, e.g. in  Ref.~\cite{Jadach:1998wp}. }.

The $\tau$ pair production process at LHC, $pp \to \tau^+\tau^- X$, can be complicated. The spectator system $X$ 
can be of a very distinct type depending on the intermediate state decaying to 
the $  \tau \tau$ pair. 
For the case of Higgs production and decay, the picture simplifies since the
Higgs is a narrow resonance and of a zero spin. The dominant production mode is $gg \to H$ fusion,
but independently of the production process it is enough to use Higgs decay products kinematics for calculation
of the decay matrix element. 
For the calculation of spin  weights in {\tt Tauola++ universal interface} or {\tt TauSpinner},
the information on four momentum of the Higgs, four momenta of $\tau$s and their decay products are sufficient. 
No approximations are involved,
independently of whether the transverse spin effects are taken  into account or not.
The production part factorizes out. Moreover the {\tt TauSpinner} algorithm is designed such that modification of the  
production mechanism, with the help of respective weights is possible. 
For the case of $2 \to 2$ processes of $\tau$ pair production,  this functionality 
is already available, see Ref.~\cite{Banerjee:2012ez} for details.

In case of the  DY process, the situation is different. Even at the lowest order
information on the flavours of the incoming quarks to the hard process is needed to introduce spin correlations.
 This defines another segment of the calculations. 
In  {\tt Tauola++ universal interface} flavours of incoming quarks are deciphered from the history in entries of
 the event record. For {\tt TauSpinner}, where we assume that the history in the event record is not available,
 four-momenta of incoming quarks are calculated from 
the virtuality and rapidity of the $\tau$ pair \cite{Czyczula:2012ny}. Then, the particular quark flavour configuration 
is chosen randomly on the basis of PDFs (e.g. ~\cite{Whalley:2005nh,LHAPDF-pdfsets}) and quark level Born cross section.
The scattering angle of the hard process is also calculated, as described in \cite{Czyczula:2012ny},
following \cite{Was:1989ce}, and is used for calculating spin weights.

The discussion above does not exhaust all details necessary for calculation of the {\tt TauSpinner } weights
for spin effects discussed in this paper, but the main 
ingredients are explained. For more details concerning the algorithms which replace the 
production matrix
elements with the help of weights, references \cite{Czyczula:2012ny,Banerjee:2012ez, TauSpinner2014}  should be consulted.
When adapting these algorithms for {\tt TauSpinner}, we could also profit from the available corresponding segments of the
 {\tt Tauola++ universal interface} and study their functionalities%
\footnote{
One can use available building blocks to evaluate the properties of the background 
without complete simulations as well. The spin correlation matrix $R_{ij}$ is calculated for 
{\tt Tauola++ universal interface} by the electroweak, one loop level, segment of the code. We have started to collect the necessary 
preliminary results in the directory  {\tt CP-tests/Z-pi}. First, we have checked
that, as in analytic form of \cite{jadach-was:1984}, only diagonal   parts of $R_{ij}$ and polarization ($R_{tz}$, $R_{zt}$) remain non zero in the 
high energy limit
(see the content of {\tt CP-tests/Z-pi/RijS-INTcosthe.root} file).
%}.
Plots for different component of $R_{ij}$  are given as a function of invariant mass and incoming quark 
flavour. The module of $R_{ij}$ is taken and integration over the hard scattering angle is performed.

Encouraged by this confirmation we have looked into the comparison of 
$R_{xx}$ and $R_{yy}$ for different values of $s$ and incoming quark flavours but as
function of the cosine of the hard scattering angle 
(see the content of {\tt CP-tests/Z-pi/Rijcostheta-S=2-6.root} and {\tt CP-tests/Z-pi/delta12.root} files).%}.

For higher values of $s$ the $R_{xx}$ and $R_{yy}$ were coinciding. 
Thus, the resulting background will feature, see Fig.~\ref{fig:Z}, transverse spin 
effects (necessary for CP Higgs measurement) as if the sample 
was not polarized, provided that the observable will symmetrize the overall 
orientation angle common for the two $\tau$ decay systems, with respect to 
the hard scattering plane.}.

\subsection{QED bremsstrahlung in decays}

So far our discussion is neglecting the  effects of bremsstrahlung in Higgs decay and final state radiation (FSR)
in DY  process. The solution implemented in {\tt TauSpinner} was discussed already in Ref.~\cite{Czyczula:2012ny}.
We have checked that test distributions used in the present paper are defined in a way which avoids potential
difficulties. All figures were reproduced for the samples where final state effects were 
included in H or $Z/\gamma^*$ decays, no differences beyond statistical fluctuations have been observed.
Also for the general case, 
it is  known that effects are small and can be factorised into a separate 
simulation block. This required discussions of the matrix element properties and we do not plan to repeat these studies here.
%This is a well known complication which need to be addressed in future. Fortunately it 
%can form separate simulation block which need to be discussed in future, when precision requirements will 
%improve. 
From the user point of view,
as discussed in \cite{Czyczula:2012ny}, to control the effect from final state bremsstrahlung, the intermediate state ($H,Z/\gamma^*$, etc.) 
decay vertex
has to be passed to the {\tt TauSpinner} as follows. FSR photons have to 
be summed into   the  four-momentum of the intermediate state $Z/\gamma^*$, but not into four momenta of $\tau^\pm$ and their decay products.
The four momentum non-conservation of such (on-fly created) vertex will be assumed to be due to FSR photons.
Detailed discussion of theoretical results behind  this solution can be found 
in the documentation of  {\tt PHOTOS} \cite{Davidson:2010ew}  and references therein.

\section {Numerical results }

Below we present benchmark numerical results for the  CP sensitive observables
discussed above. 
We show, that the effects of transverse spin correlations, already published in \cite{Davidson:2010rw} using {\tt Tauola++ universal interface} 
and obtained also by other authors with independent approaches, can be reproduced using {\tt TauSpinner}.

\subsection {Case of the 125 GeV Higgs }
\label{sec:Case125}

The acollinearity angle in the $H \to \tau \tau$ with both  $\tau^\pm \to \pi^{\pm} \nu$, which requires 
precise reconstruction of the Higgs rest frame, can be realized  in the muon collider
 but  was considered difficult
in LC or LHC experiments, due to missing neutrino momenta from $\tau$ decays.
Nonetheless, we show this observable in Fig.~\ref{fig:acoll} and included it into our benchmarks discussed in 
Appendix~\ref{sec:thetest} because of its straightforward theoretical interpretation. 
Because of the simple kinematical constraints, in the rest frame of the Higgs boson, $\tau$'s are back-to-back 
and acoplanarity angle between decay products $\pi^+ \pi^-$ is above 2.8. 
When zooming into this region (right plot of Fig.~\ref{fig:acoll}), a clear difference between spectra
for scalar and pseudoscalar Higgs boson is observed. However, to measure it experimentally, very precise reconstruction
of the Higgs boson rest frame would be mandatory.

\begin{figure}[htp!]
\begin{tabular}{ccc}
  \includegraphics[width=0.48\columnwidth]{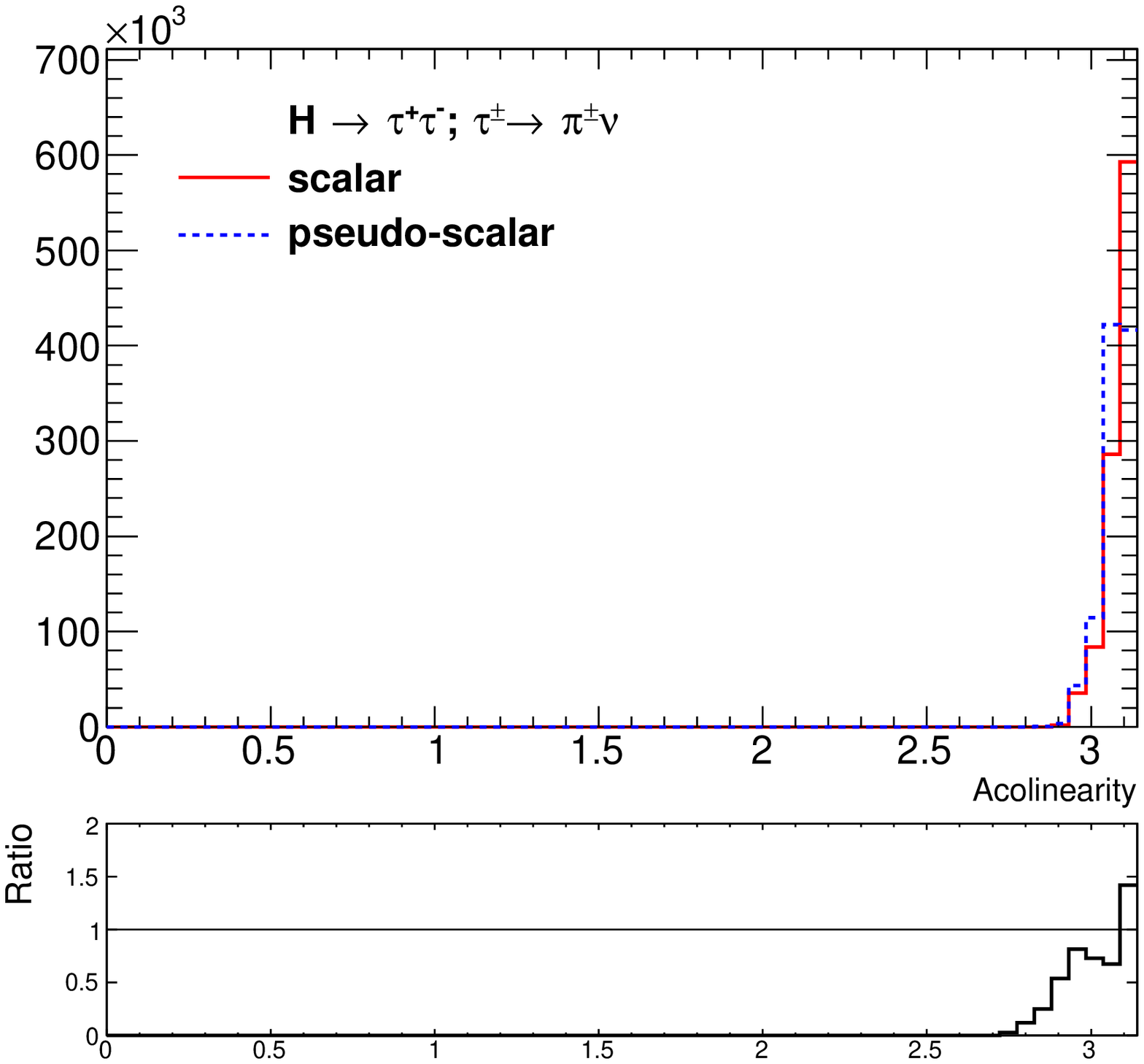} &
  \includegraphics[width=0.48\columnwidth]{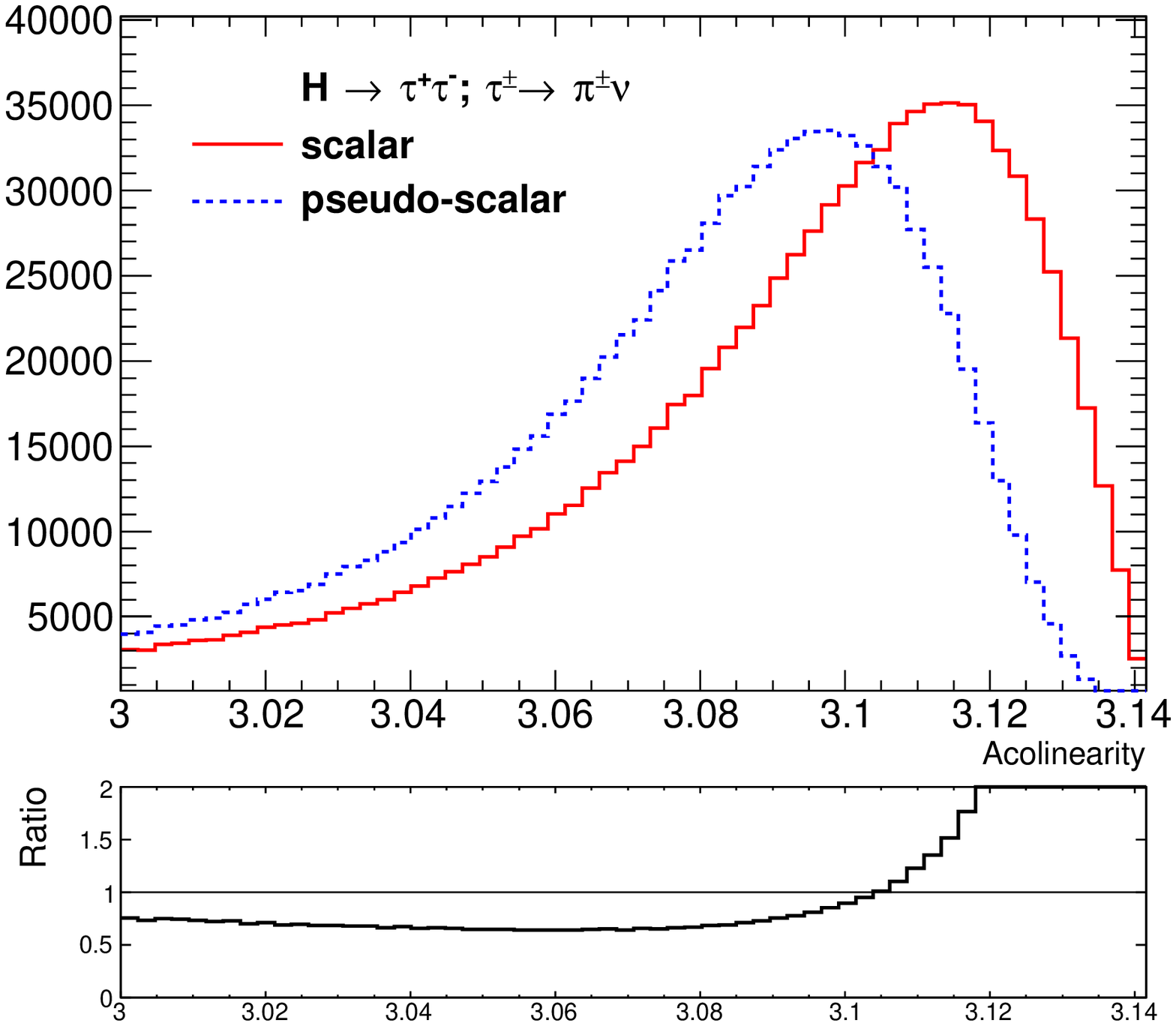} 
\end{tabular}
\caption{
The acollinearity distributions of the $\pi^+ \pi^-$ in $H \to \tau^+ \tau^- $, $\tau^\pm \to \pi^\pm \nu$ decays. 
Red line corresponds to the scalar and blue dashed line to the pseudoscalar case. 
Black line shows the ratio of the distributions.
On the right side, the acoplanarity range is zoomed 
to show  its endpoint region. 
\label{fig:acoll}}
\end{figure}

We consider the $H \to \tau^+ \tau^-$, $\tau^\pm \to \rho^\pm \nu$,
$\rho^\pm \to \pi^\pm \pi^0$ decay chain and one-dimensional angular distribution of the acoplanarity 
angle $\theta_{acop}$ between the $\pi^+ \pi^0$  and the $\pi^- \pi^0$ decay planes in the $\rho^+ \rho^-$ pair 
rest-frame (Ref.~\cite{Bower:2002zx}).
 In view of the Standard Model Higgs boson of a mass of 125 GeV,
it is a  good candidate for a CP sensitive observable at LHC. 
This distribution is CP sensitive if events are separated%
\footnote{This separation, thanks to the properties of $\tau$ decay matrix elements, correlates the $\rho$ decay plane
with the one spanned on $\tau$ and $\nu_\tau$ momenta.} 
into two categories, accordingly to the sign of the product $y_+ \cdot y_- $. 

We recall this distribution\footnote{Details on the program used in preparation of this plot are given in Appendix 
\ref{sec:thetest}.} in Fig.~\ref{fig:rhorho}.  We split events into two separate categories and compare the 
case of the scalar (red) and the mixed scalar-pseudoscalar state (blue dashed) with assumed mixing angle 
$\theta=0.2$. The  $\theta_{acop}$ distribution relies on measurement of the four-momenta (in the laboratory frame)   of the 
$\pi^+ \pi^0$ and $\pi^- \pi^0$ only. No reconstruction of the rest frames of Higgs and/or $\tau^\pm $ is required.
Obviously, the nature of the observable is at least three  dimensional
and the multi-dimensional fit in the space $(\theta_{acop},  y_-, y_+ )$  offers 
additional increase in sensitivity of this measurement.

Fig.~\ref{fig:acopRF} shows the same distribution but when  $y_\pm$ energy differences 
are calculated in  $\tau^\pm$ rest frames instead of the laboratory frame. 
Reconstruction of $\tau^\pm$ rest frames can be in part achieved with 
the help of $\tau$ decay vertex measurement. 
Comparison of Fig.~\ref{fig:acopRF} and~Fig.~\ref{fig:rhorho} allows to estimate available potential improvement
which may be accessible also at LHC. %as at LC.

This optimization can be realized in part,
if the  $\tau$ decay vertex is  measured to a limited precision, as 
used in  \cite{Desch:2003mw}.  This helps resolving twofold ambiguity in reconstructing the $\tau$ momentum.
Other examples where the reconstruction of the decay vertex is used
as an essential part of the observable, are given in  
\cite{Rouge:2005iy} and  also\footnote{
Note that the angles $0<\phi^*< \pi $ and $0<\Psi_{CP}^* < \pi$ of Ref.~\cite{Berge:2008dr} 
are  related to  our angle $0<\theta_{acop} < 2 \pi$. The  $\phi^*=\arccos(\cos (\theta_{acop}) )$ and 
$\Psi_{CP}^*=\arcsin (\sin (\theta_{acop}) )$. Different, but correlated, physics input is used for the definition of planes.
In Ref.~\cite{Desch:2003rw} planes spanned on momenta  $\pi^\pm$ and $\pi^0$  from $\rho^\pm$ decays are used.  
In ~\cite{Berge:2008dr} the plane spanned by $\tau$ direction and its charged decay product is used 
instead; exactly as the angle $\phi^*$ of Ref.~\cite{Kramer:1994jn}. 
The actual choice will depend on how well measured are $\pi^0$ directions and precision of reconstruction
of the $\tau$ decay vertex. If all these quantities are to be shown to be measurable, one should 
consider simultaneous fit to multidimensional distribution over  all of the variables.} in
 Ref.~\cite{Berge:2008dr}. Naturally, extension to other decay modes of $\tau$'s, 
especially to 3 $\pi$ channels will be beneficial as well. These channels offer
substantial increase of the sample statistics, and also, if ignored would contribute 
to the background. Many options, with optimal choices depending  on 
details of detector 
response, the hard scattering  and $\tau$ decay distributions, can be envisaged.
%Such discussions however must be performed in the context of the detector response.
%Careful study of all possible detector information is needed.

\begin{figure}[htp!]
\begin{tabular}{ccc}
  \includegraphics[width=0.48\columnwidth]{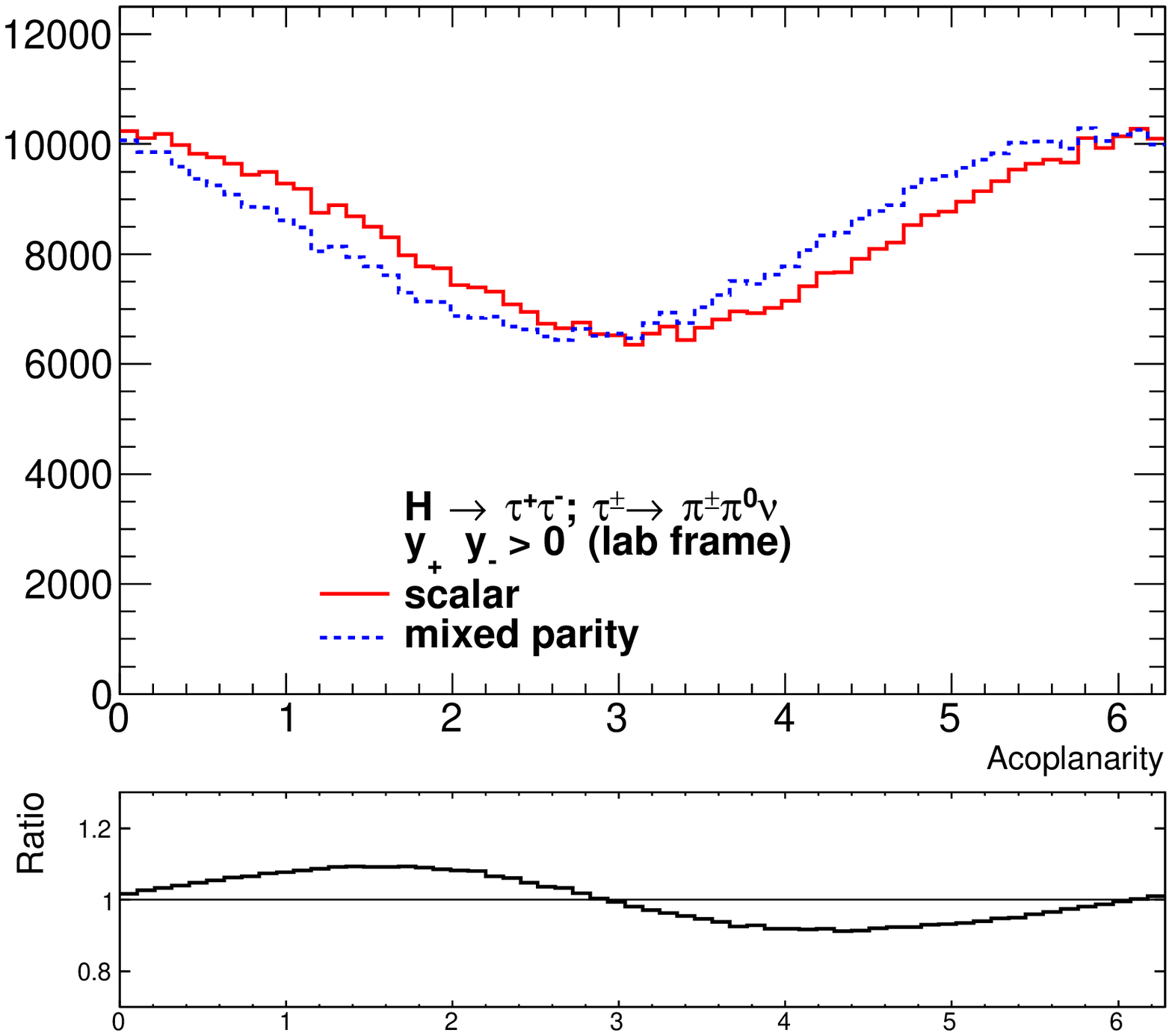} &
  \includegraphics[width=0.48\columnwidth]{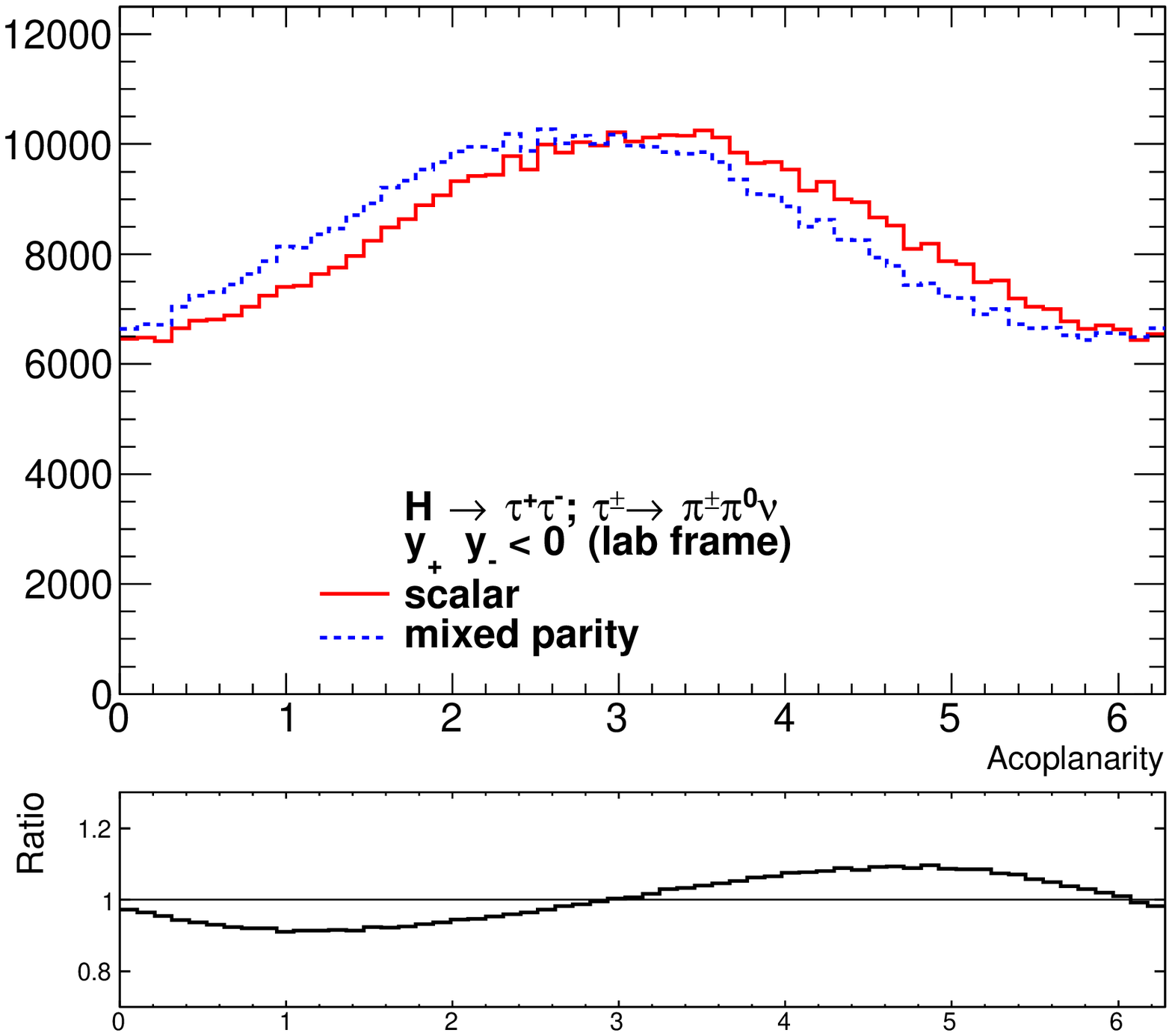} 
\end{tabular}
\caption{
The acoplanarity distribution for the $H \to \tau^+ \tau^- $, $\tau^\pm \to \pi^\pm \pi^0 \nu$ decays
(see text for more details). Left plot for events with  $y_- \cdot y_+ >0$, right plot for  $y_- \cdot y_+ <0$.
Compared are scalar (red) and mixed scalar-pseudoscalar (blue dashed), with mixing angle $\theta=0.2$, cases.
The  $y_\pm$ variables are calculated in the laboratory frame.
\label{fig:rhorho}}
%\end{figure}
%\begin{figure}[htp!]
\begin{tabular}{ccc}
  \includegraphics[width=0.48\columnwidth]{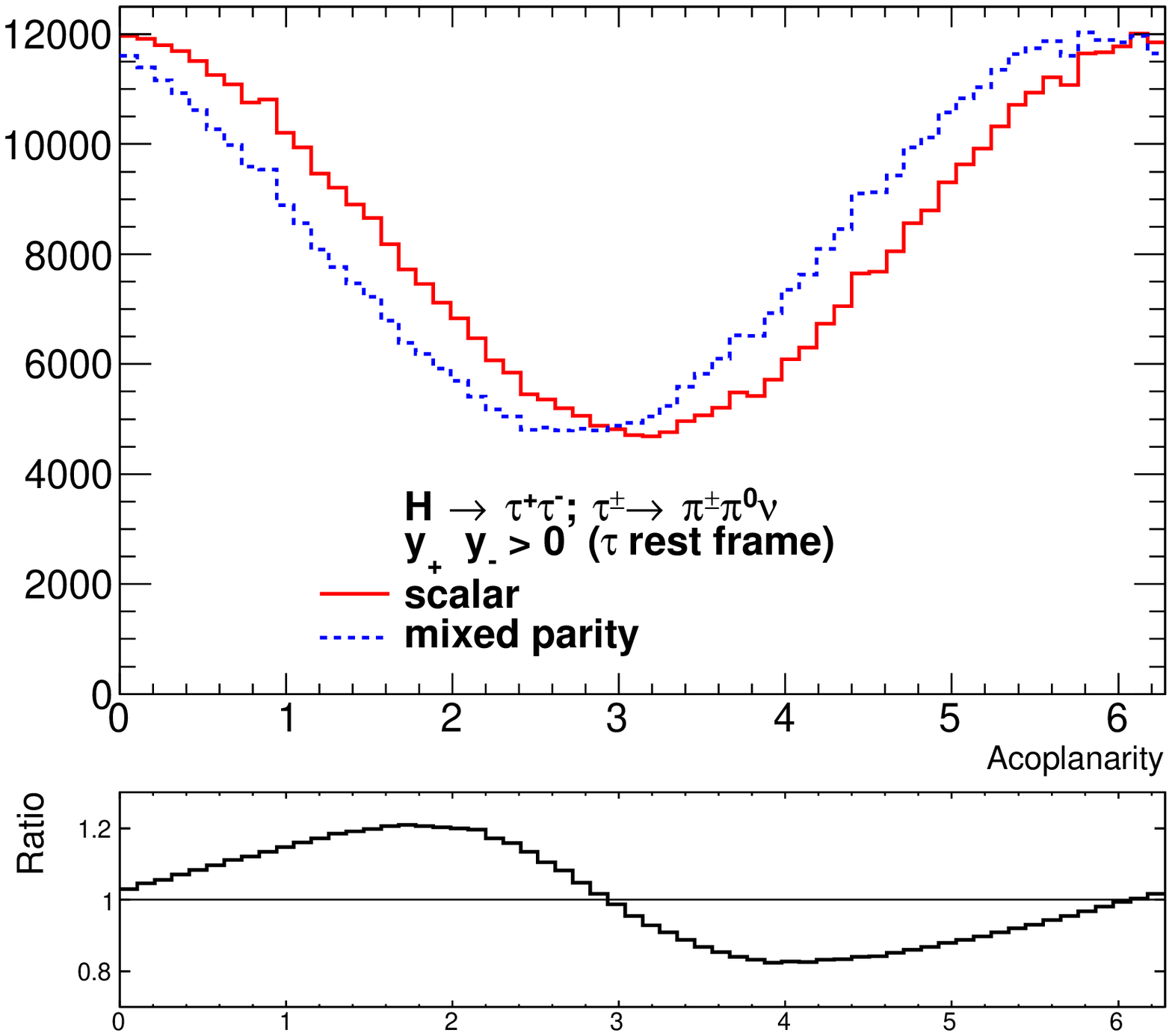} &
  \includegraphics[width=0.48\columnwidth]{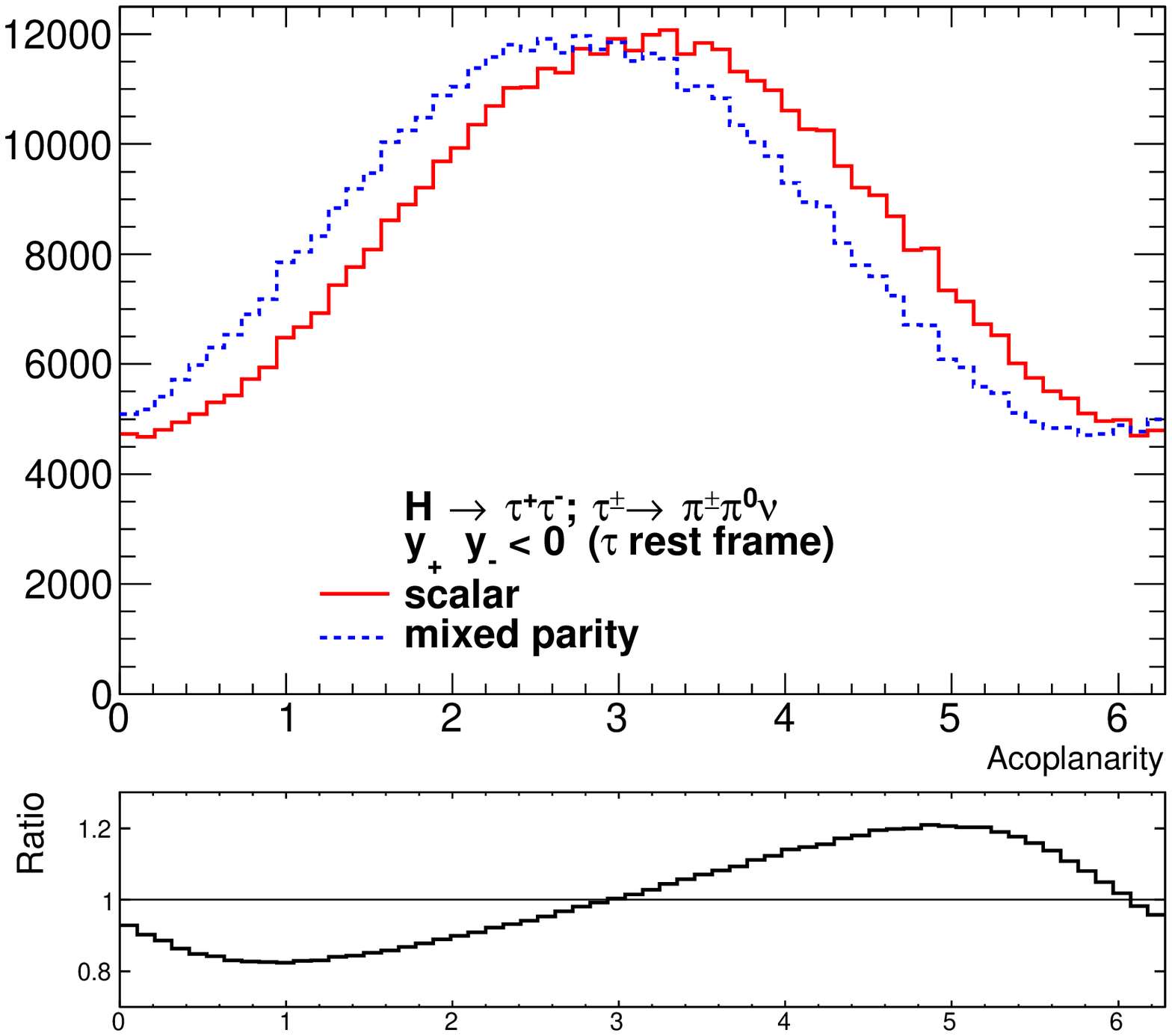} 
\end{tabular}
\caption{
The same as Fig.~\ref{fig:rhorho} but with $y_\pm$ variables
calculated in  $\tau^\pm$ rest frames.
\label{fig:acopRF}}
\end{figure}

\subsection {Case of Drell-Yan background}
\label{sec:CaseDY}

Studies of the transverse spin effects and of CP sensitive observables can not 
be completed for the Higgs boson unless properties of the main background, 
that is DY $Z/\gamma \to \tau^+ \tau^-$, are studied as well.  At LHC, this dominant background 
is difficult to  separate from the Higgs boson signature~\cite{ATLASHtautau}.
Both Atlas and CMS use embedded $\tau$ samples for the estimation of the background
\cite{ATLASHtautau, CMSHTautau}.
Control of the transverse spin effects can be realized in this case with the help of
weights of the new version of {\tt TauSpinner} presented in this paper.

\begin{figure}[htp!]
\begin{tabular}{ccc}
  \includegraphics[width=0.48\columnwidth]{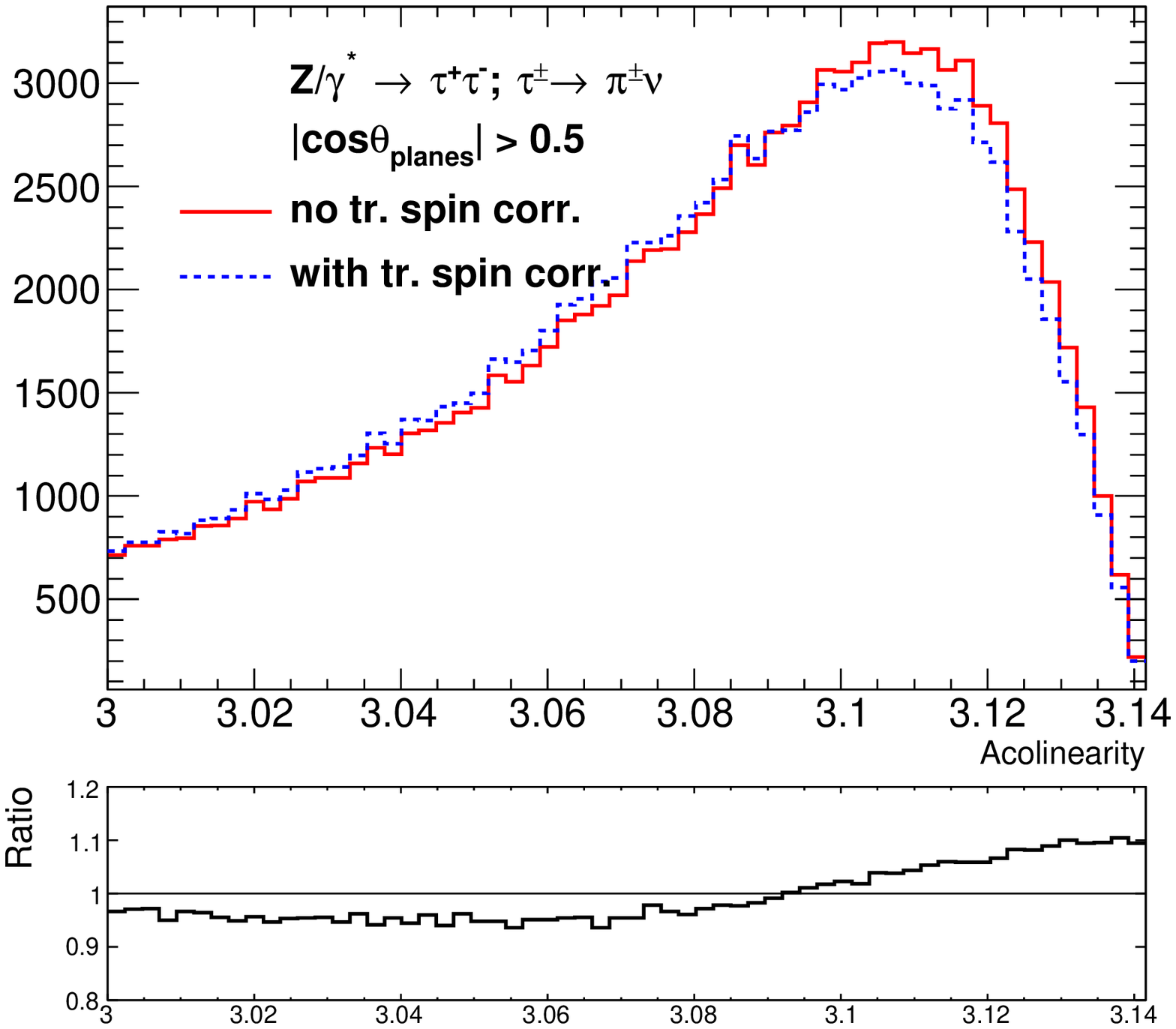} &
  \includegraphics[width=0.48\columnwidth]{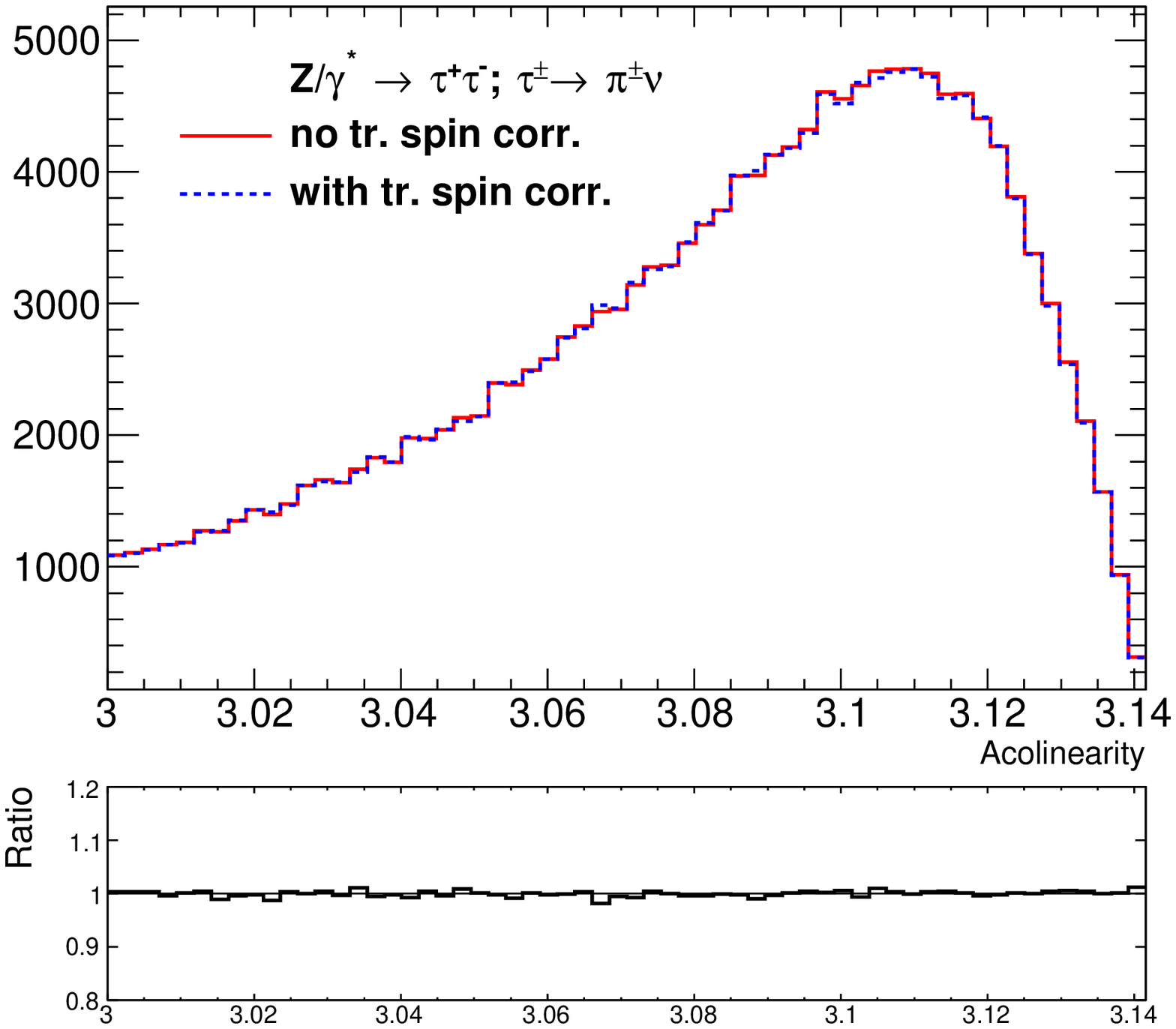} 
\end{tabular}
\caption{
The acollinearity of $\pi^+\pi^-$ directions in the rest frame of intermediate $Z/\gamma^*$ state, invariant mass of the $\tau^+\tau^-$ pair is requested to be larger than 60 GeV. 
An additional cut has been applied (left plot), as defined in the text, to enhance effect
from transverse spin correlations. Without this cut (right plot) effect
is substantially smaller than statistical fluctuations.
\label{fig:Z}}
\end{figure}

Numerical results for the DY  sample  with $m_{\tau \tau} > 60$ GeV  are shown in Fig.~\ref{fig:Z},
for the acoplanarity of the $\pi^+\pi^-$ directions, i.e.  the same variable as in Fig.~\ref{fig:acoll} for the Higgs boson. 
Note that because of the line-shape of the $Z/\gamma^*$, i.e. large spread of the $\tau^+\tau^-$ virtualities, the 
end-point of the $\pi^-\pi^+ $ acollinearity distribution is not as sharp 
as for the Higgs boson. For DY events, 
no transverse spin effects can be seen (right plot).  To nevertheless enhance effects from the spin correlations,
an additional cut on $|\cos \theta_{planes}|> 0.5$ is introduced (left plot), leading to a difference of  
20\% in the shape of the falling edge.
The $\theta_{planes}$ is an angle between planes defined by: $\tau^\pm$ momentum -- beam
momentum (first plane) and  $\tau^\pm$ momentum 
and its decay product $\pi^\pm$ momentum (second plane). 

The DY background is also characterized by a simply flat distribution of the acoplanarity angle,  
in case of   $Z \to \tau^+ \tau^-$, $\tau^\pm \to \rho^\pm \nu$ decay chain; 
where  $\theta_{acop}$ is defined between $\pi^+ \pi^0$  and $\pi^- \pi^0$ decay 
planes in the $\rho^+ \rho^-$ pair rest frame. The distribution, shown in Fig.~\ref{fig:acopRF} for the Higgs case,
becomes a sensitive observable only if dependent on the sign of  $ y_- \cdot y_+$.
This is not the case of the Z boson (independent of the sign of $ y_- \cdot y_+$) and this feature may be used to control 
the background, unless (unlikely) destroyed accidentally by some selection cuts.

We have cross-checked these results with samples generated including transverse-spin correlations directly (i.e.
using {\tt Tauola++ universal interface}). In this case, the transverse spin effects for the DY  production are activated 
simultaneously with the electroweak loop corrections, see \cite{Davidson:2010rw}. 

\section { Summary}
\label{sec:summary}

In this paper, we have presented a new extension of the {\tt TauSpinner} package, 
namely the implementation of the transverse spin correlations in the decay of 
scalar/pseudoscalar state into a pair of $\tau$ leptons as well 
as for DY  $Z/\gamma^*$ processes.
The {\it post-fact} modeling of the correlations is achieved by calculating 
weights on the event-by-event basis using only information on the kinematics
of the outgoing $\tau$ decay products. This provides a  convenient tool,
where multiple models can be studied on the correlated events without the need 
for CPU-intensive simulations. The question of weights to model/replace 
matrix elements used for generating scalar/pseudoscalar state production is postponed to a forthcoming 
paper.
 
The new functionality has been achieved by porting and adapting code 
developed for the {\tt Tauola++ universal interface} package, so it can be used 
on already generated events where such correlations were included or not.

We have recalled two observables sensitive to the transverse spin correlations 
in case of $H \to \tau \tau$ decays, which have been proposed already 
some time ago: acolinearity  in case of $\tau \to \pi^\pm$ decay, 
and acoplanarity  in case of $\tau \to \pi^\pm \pi^0 \nu$ decays. 
They have been used to provide numerical benchmark
of this implementation, for the Higgs mass of 125 GeV, as recently 
discovered by the LHC experiments.

We have also discussed transverse spin correlations for the DY  $Z/\gamma^*$ background process,
which can now be modeled 
with the  {\tt TauSpinner} package.
We  have shown 
that the effect for  background is marginal (as expected). Nevertheless, given the fact that
experimental selection may enhance such effects it is considered as important
to have it also available.

We have not addressed here the question of the feasibility to  observe transverse 
spin correlations in the analyses of $H \to \tau \tau$ channel at LHC. 
Given nowdays very refined analysis techniques, even rather small effects
can be turned into successful measurements. Our aim was to prepare, 
describe and validate a tool which may be useful for such a goal.  
 
We have concentrated on the Higgs decays. One expects that for the distinct parity of the Higgs-like object,
its production hard scattering matrix element should differ as well.
Respective modifications to the production matrix elements of the user choice
(and not only to the spin correlation weight discussed in the present paper) 
has been  implemented  for the $2 \to 2$ hard processes
since some time now~\cite{Banerjee:2012ez}. 
Our intention is to extend this functionality also to the 
vector boson fusion production processes \cite{Kalinowski:2014}, 
that is  kinematical configurations
of final states consisting of $\tau^+\tau^-$ pair and two jets. 
Implementation of the transverse spin effects, for such user provided 
non-standard physics processes, may be considered in the future.

\section*{Acknowledgements}
%Discussions with ...are acknowledged.

This research was supported in part by
the Research Executive Agency (REA) of the European Union under
the Grant Agreement PITN–GA–2012–316704 (HiggsTools) and by funds of Polish National Science
Centre under decision DEC-2011/03/B/ST2/00107.

\providecommand{\href}[2]{#2}
%\bibliography{Tauola_interface_design}{}
%\bibliographystyle{utphys_spires}

\appendix
\newpage
\section{TauSpinner technical details}
\label{sec:technical}

\subsection{Initialization}

In case of transverse spin effects, the following changes have to be introduced 
for {\tt TauSpinner} initialization. The rest of its execution is performed as explained in the 
previous publications \cite{Czyczula:2012ny,Banerjee:2012ez}. In particular, all weights will still be calculated 
by calling the function\\
 {\tt calculateWeightFromParticlesH}\footnote{For the purpose of adding new functionality, {\tt TauSpinner} must access
private fields and functions of {\tt TauolaParticlePair} class from {\tt Tauola++ universal interface} library.
This has been resolved through a proper {\tt friend} declaration in {\tt TauolaParticlePair.h} header file. No other modifications were needed.}.

Let us give a few details concerning different options:

{\bf Case of Higgs bosons}
\begin{itemize}
\item
At initialization stage: \\
             {\tt setHiggsParametersTR(-1.0, 1.0, 0.0, 0.0); } for scalar Higgs or \\
             {\tt setHiggsParametersTR( 1.0,-1.0, 0.0, 0.0); } for pseudo-scalar Higgs 
\item
For mixed parity state, use: \\
             {\tt double theta = 0.2; } \\
             {\tt setHiggsParametersTR(-cos(2*theta),cos(2*theta), }\\
             {\tt -sin(2*theta),-sin(2*theta)); }
\end{itemize}

{ \bf Case of DY process%
    \footnote{Transverse spin correlations for DY are  available  with the development 
    version or release version 
    distributions of {\tt Tauola++ v1.1.5} or later.}}
\begin{itemize}
\item
At initialization stage: \\
    {\tt setZgamMultipliersTR(1.0,1.0, 1.0, 1.0);} \\
    the $R_{xx}, R_{yy}, R_{xy} $, and $R_{yx} $ components of the 
    density matrix will be multiplied by these coefficients. 
    The two files named {\tt  table1-1.txt, table2-2.txt }  
    have to be present in the directory  of the executable main 
    program, exactly as in the case of {\tt Tauola++ universal interface} \cite{Davidson:2010rw}.
    If the  tables are absent, or if their name is distinct, 
    the transverse components of $R_{ij}$ will be equal to zero.    
\item
At execution stage: \\
The method {\tt getZgamParametersTR(RXX, RYY, RXY, RYX);  } 
can be used to port the numerical values of 
$R_{xx}, R_{yy}$, $R_{xy}, R_{yx}$ to the user program  to monitor the values or to modify them
by repeating the following: \\
{\tt setZgamMultipliersTR(...);} \\
{\tt WT1 = calculateWeightFromParticlesH(...);} \\ 
for each individual event.
Motivated by the results of our tests,  
we have chosen  $R_{xy}$ and $R_{yx}$  to be zero in all DY
cases.
\end{itemize}

Further details are given in the following sections and in {\tt README} files of the distribution tar-ball.
All the plots of the present paper can be reproduced with the help of new demonstration programs 
which are included in the {\tt Tauola++ v1.1.5} (or later) distribution tar-ball.
We are using the {\tt ROOT}-based {\tt MC-TESTER} package \cite{Davidson:2008ma} to facilitate histograming and plotting.

\subsection{Electroweak corrections}
\label{sec:transverse-spin-effects}
In  {\tt Tauola++ universal interface} transverse spin effects   are activated 
together with electroweak corrections of the {\tt SANC} library, see Ref.~\cite{Andonov:2008ga}. Now, this is the case in {\tt TauSpinner} as well.
The  $R_{xx}$ and $ R_{yy}$ components of $\tau^+ \tau^-$ pair spin density matrix have been
ported to the {\tt TauSpinner} code. Other, non-diagonal  
components, have been demonstrated to be suppressed to, or below the few 
percent level and are not taken into account. See the plots of the  file
{\tt CP-tests/Z-pi/RijS-INTcosthe.root} discussed in the main body of the paper.

\subsection{Installation}
\label{sec:installation}

As  auxiliary material to this paper, the directory {\tt TAUOLA/TauSpinner/examples/CP-tests}
has been prepared. The main program {\tt CP-test.cxx} is based on
the default example program {\tt tau-reweight-test.cxx} located in the {\tt TAUOLA/TauSpinner/examples}
directory. In order to compile it one has to:

\begin{itemize}
\item Configure {\tt Tauola++} for compilation with {\tt TauSpinner},
      {\tt HepMC} \cite{Dobbs:2001ck} and {\tt MC-TESTER} \cite{Davidson:2008ma}.
      See \cite{Davidson:2010rw} and \cite{Czyczula:2012ny} for instructions on how to define the appropriate paths.
\item If using {\tt Tauola++} distribution tarball for LCG%
      \footnote{With fixed initialization for $\tau$ decay matrix elements
                and prepared for installation in LCG library~\cite{LCG}.},
      available on the project website \cite{tauolapp:website}, no additional
      configuration is required. In case of complete distribution%
      \footnote{Enabling, in particular, the change of $\tau$ decay matrix elements.},
      execute {\tt ./configure} with appropriate paths in the {\tt TauSpinner/examples} directory.
\item Execute {\tt make} in {\tt CP-tests}.
\end{itemize}

The example located in this directory is prepared to work without  modifications.
Please see comments in the code for possible options.

\subsection{Executing tests}
\label{sec:thetest}
The directory {\tt CP-tests} contains four tests located in subdirectories
{\tt H-rho}, {\tt H-pi}, {\tt Z-rho} and {\tt Z-pi}. Each test
can be executed by the command {\tt ../CP-test.exe} in one of those subdirectories.
Each subdirectory contains a small {\tt HepMC} file {\tt events.dat} with a sample of
100 events to test the installation. 
See Appendix \ref{sec:generating_datafiles} for instructions on how to set up generation
of larger data samples using {\tt Tauola++} and {\tt Pythia8}.
Executing the program with no parameters 
will process the predefined {\tt events.dat} sample.
Any   sample can be processed with: \\
\\
{\tt ../CP-test.exe <data\_sample> [<optional\_events\_limit>]} \\
\\
For each  subdirectory, an appropriate {\tt MC-TESTER}
user analysis script is provided. Appropriate parameters are set with the help of  
{\tt setHiggsParametersTR} and {\tt setZgamMultipliersTR}
 in the main program {\tt CP-test.cxx} file.
Note that subdirectories {\tt Z-rho} and {\tt Z-pi} contain previously generated  tables of electroweak one loop level 
results for the quark level differential cross section and $R_{ij}$%
\footnote{ See {\tt Z-pi/table1-1.txt, Z-pi/table2-2.txt}. Subdirectory {\tt Z-rho} contains
symbolic links to these two files. See \cite{Davidson:2010rw} for instructions on how to
generate these tables with modified parametrization.},
which are needed for these tests (see Appendix \ref{sec:transverse-spin-effects}).

For  tests (and for our figures), {\tt MC-TESTER} data files have been used with the results
of processing samples of 1M~events for pp@8TeV (or @14TeV) collisions generated with {\tt Pythia8} \cite{Sjostrand:2007gs}
hard process option {\tt HiggsSM:ffbar2H} ( {\tt WeakSingleBoson:ffbar2gmZ}) turned on.
We have checked that  variation of the Higgs mass in range 120-125 GeV does not affect results beyond the statistical fluctuations.
This applies to the case when ISR, FSR is activated in {\tt Pythia}
as well\footnote{ In this case, a large number  of final states can be recognized by {\tt MC-TESTER} -- result of FSR activity in Z or H decays.
On some platforms, for bigger samples, this may cause buffer overflow problems.
Our examples are not expected to work for all 
possible options of constructing and storing the event records.
They are supposed to work for the
properly prepared data files only.}.
We have used {\tt LHAPDF} dataset {\tt cteq6ll.LHpdf} \cite{LHAPDF-pdfsets}.
{\tt MC-TESTER} user analysis scripts are adapted to the appropriate Higgs ($Z/\gamma^*$) states.

\subsection{Analyzing results}

The result of the test described in previous section is stored in {\tt mc-tester.root} file.
It contains all plots defined in user analysis script file%
\footnote{User analysis {\tt ROOT} scripts are the {\tt .C} files located
in the directory where the test is run. The name of those files ends with
{\tt *UserTreeAnalysis.C}). Plot definitions located in these files
can be modified as needed, however new plot definitions mean that
benchmark files provided with the distribution cannot be used for comparison.
Keep in mind that in order for {\tt MC-TESTER} to work correctly {\tt \$(MCTESTERLOCATION)} environmental
variable must be set.}.
To compare two {\tt MC-TESTER} rootfiles use {\tt compare.sh}  located in the {\tt CP-tests} directory: \\
\\
{\tt ./compare.sh <file1.root> <file2.root>} \\
\\
The benchmark distributions will be stored in the  section {\tt USER HISTOGRAMS}
of the booklet {\tt tester.pdf} produced during comparison.
Each directory provides one or more benchmark files that can be used if
no changes have been introduced to user analysis script files.

Executing the program in the directory {\tt CP-tests/H-rho} generates Figure \ref{fig:rhorho} 
if root files {\tt scalar.root, thet-0.2.root} are used as inputs.
With {\tt scalar-tauframe.root, thet-0.2-tauframe.root} used as inputs, the plots for Fig.~\ref{fig:acopRF} are generated instead.
These plots, to a large degree, coincide with  Figure 2 of Ref.~\cite{Desch:2003rw}.
Note the different choice of quantization frames used in these papers. 
If the directory {\tt CP-tests/H-pi} is used, Fig.~\ref{fig:acoll} is generated,
reproducing  Fig.~3 from \cite{Kramer:1994jn} (files {\tt scalar.root, pseudoscalar.root} are  used). 
Not only the acollinearity
of $\pi^\pm$ directions in the rest frame of the Higgs is histogrammed, but also the  
acoplanarity angle for the $\pi^- \tau^-$ and $\pi^+ \tau^+$ planes
of the same frame  (it is not included in our paper).

Numerical results for the DY sample of 1M~events are collected  in the files:
{\tt start60.root} and {\tt transverse60.root} in the {\tt CP-tests/Z-pi} subdirectory,
and were used for the plot shown in Fig.~\ref{fig:Z} for the case with  
the selection $|\cos \theta_{planes}|> 0.5$,
and files {\tt start60-noc.root} and {\tt transverse60-noc.root} without this selection.

\subsection{Generating data files}
\label{sec:generating_datafiles}

In this section we describe how data files can be generated for the tests.
For this purpose, a subdirectory {\tt CP-tests/generate-datafiles} has been
provided with a program specifically configured for generating data files for
these tests. Note that using this subdirectory requires a path to {\tt Pythia8}
to be provided during the configuration  of {\tt Tauola++} in addition to other
paths%
\footnote{Contrary to {\tt TAUOLA/examples} subdirectory,
this program is set up only for {\tt pythia} version 8.180 or later. When  using older versions,
changes to the use of {\tt pythia8 HepMC} interface may have to be introduced.}.
To compile this program, execute {\tt ``make generate''} in the {\tt CP-tests}
directory. To run it, execute:\\
\\
{\tt ./generate.exe <output\_file> <pythia8\_config> <decay\_channel> <events\_count>} \\
\\
Two {\tt Pythia8} config files have been prepared: {\tt pythia.Z.conf}, {\tt pythia.H.conf},
and they can be used in combination with {\tt Tauola++} decay channel 3 or 4 to provide
samples for $\pi$ or $\rho$ respectively%
\footnote{Tests described in this paper have been configured for proton-proton collisions
at $\sqrt{s} = 14 TeV$, therefore {\tt Pythia8} configuration files use the same setup.}.

The program is set up to strip event of all particles other than $H$ or $Z$ boson,
$\tau$ and $\tau$ decay products. This information is enough for {\tt TauSpinner}
to work and output files are relatively small (around 4GB per 1M~events).

\end{document}